\documentclass[twoside,a4paper,preprint,11pt]{article}

\usepackage{blindtext}

\usepackage[preprint]{irrj2e}
\usepackage{graphicx}
\usepackage{amsmath}
\usepackage{natbib}
\usepackage{amsfonts}
\usepackage{booktabs}
\usepackage{float}
\DeclareMathOperator*{\argmax}{arg\,max}

\usepackage{lastpage}

\irrjheading{1}{2025}{1-\pageref{LastPage}}{2/25}{-}{10.54195/irrj.00000}

\ShortHeadings{Better Projections Improve ColBERT}{Clavié et al.}
\firstpageno{1}

\begin{document}

\title{Simple Projection Variants Improve ColBERT Performance}

\author{\name Benjamin Clavié \email ben@mixedbread.com | bc@nii.ac.jp \\
       \addr Mixedbread AI and National Institute of Informatics (NII)
       \AND
       \name Sean Lee \\
       \addr Mixedbread AI
       \AND
       \name Rikiya Takehi \\
       \addr Mixedbread AI and Waseda University
       \AND
       \name Aamir Shakir \\
       \addr Mixedbread AI
       \AND
       \name Makoto P. Kato \\
       \addr University of Tsukuba and National Institute of Informatics (NII)}

\editor{My editor}

\maketitle

\begin{abstract}
Multi-vector dense retrieval methods like ColBERT systematically use a single-layer linear projection to reduce the dimensionality of individual vectors. 

In this study, we explore the implications of the MaxSim operator on the gradient flows of the training of multi-vector models and show that such a simple linear projection has inherent, if non-critical, limitations in this setting. We then discuss the theoretical improvements that could result from  replacing this single-layer projection with well-studied alternative feedforward linear networks (FFN), such as deeper, non-linear FFN blocks, GLU blocks, and skip-connections, could alleviate these limitations.

Through the design and systematic evaluation of alternate projection blocks, we show that better-designed final projections positively impact the downstream performance of ColBERT models. We highlight that many projection variants outperform the original linear projections, with the best-performing variants increasing average performance on a range of retrieval benchmarks across domains by over 2 NDCG@10 points.
We then conduct further exploration on the individual parameters of these projections block in order to understand what drives this empirical performance, highlighting the particular importance of upscaled intermediate projections and residual connections. As part of these ablation studies, we show that numerous suboptimal projection variants still outperform the traditional single-layer projection across multiple benchmarks, confirming our hypothesis.

Finally, we observe that this effect is consistent across random seeds, further confirming that replacing the linear layer of ColBERT models is a robust, drop-in upgrade. 
\end{abstract}


\section{Introduction}

During the past several years, a rapidly growing subfield of Information Retrieval (IR) has been Neural IR, largely consisting of deep learning methods built on the Transformer architecture~\citep{transformers} and leveraging the pretrained weights of language models such as BERT~\citep{bert}. Within Neural IR, many individual paradigms have appeared. Among others, single-vector dense retrieval~\citep{dpr}, learned sparse retrieval~\citep{splade,learnedsparse} and late-interaction multi-vector retrieval, frequently referred to as ColBERT~\citep{colbert} after the model introducing this paradigm, have been particularly notable.

Multi-vector models, e.g. ColBERT and its variants, work by encoding both queries and documents into many small token-level vectors, where the outputs of the fine-tuned backbone models are passed through a linear projection to lower their dimensions, in contrast to single-vector methods where the original model dimensions are frequently used~\citep{dpr,e5}. These representations are then subsequently used at retrieval-time to compute fine-grained interactions between documents, using the MaxSim operator, further explained in Section~\ref{sec:maxsim}. 

Currently, all existing multi-vector models largely follow variations of the original ColBERT architecture, and tweaks have largely focused on the use of different backbone models to unlock novel modalities or context length capabilities~\citep{colpali,reasoncolbert}, better training methods~\citep{jacolbertv2.5,colbertv2} or the introduction of modality-specific components~\citep{videocolbert}, with all models adopting a form of the original architecture, passing the final backbone model's hidden states to a single-layer linear projection to obtain the final output representations.

In parallel, much work in deep learning has focused on further our understanding and improving the architecture of neural models~\citep{dlimprov1,dlimprov2}. A large stream of this work has been the exploration of the impact of the feedforward block~\citep{notallyouneed,ffnmemory}, of which ColBERT's single-layer linear projection is the simplest form, as well as how to design better ones~\citep{glu,glushazeer,silu,swish}. While not immediately applicable, recent work in information retrieval exploring the use of hypernetwork as document encoders further informs this point, showing that better feedforward layer construction results in better relevance scoring~\citep{hypencoder}. However, there currently has been no in-depth study evaluating the impact, or lack thereof, of ColBERT's final projection on downstream performance.

\subsection{Contributions}

In this paper, we seek to explore whether or not different projection heads could result in greater downstream ColBERT performance. 

We first highlight the impact that the MaxSim operator has on gradient flow, before discussing the potential limits of single-layer projections which can be further compounded by this unique gradient flow. 

Building on the existing deep learning literature, we then propose a series of modifications to the final feedforward block of ColBERT models and demonstrate how their properties could result in improved retrieval performance. 

We empirically demonstrate the correctness of our hypothesis, showing that improved projection heads consistently outperform the widely-used single-layer projection on all evaluated benchmarks, with the best-performing variant improving overall performance by over 2NDCG@10 points when averaged over multiple common benchmarks.

Finally, we explore individual factors contributing to this improved performance and further demonstrate that projection head design matters, with certain ``improvements'' resulting in worsened performance, likely due to the conflicting theoretical properties manifest during training. 

\section{Related Work: Improving Multi-vector Retrieval}

Since the original release of ColBERT, substantial work has focused on improving the retrieval performance of multi-vector models, and extending them for additional uses. Notably among those, ColBERTv2~\citep{colbertv2} introduced significant performance gains by leveraging knowledge distillation over a large number of teacher-scored documents for each query. JaColBERTv2.5~\citep{jacolbertv2.5} and Jina-Colbert-v2~\citep{jinacolbertv2} subsequently introduced further refinement to the training process, showing strong empirical downstream gains. 

In the meantime, the multi-vector retrieval paradigm has been demonstrated to be easily transposable to various domains, reaching strong multilingual~\citep{colbertxm,jinacolbertv2}, cross-lingual~\citep{colbertx} results, but also modalities, with multi-vector models reaching state-of-the-art performance in text$\rightarrow$image~\citep{colpali} and text$\rightarrow$video~\citep{videocolbert} retrieval without any major changes to the underlying late-interaction mechanism. Further explorations into multi-vector multimodal retrievers have recently highlighted remarkable parameter-efficiency, with 300 million parameters multi-vector retrievers reaching error reduction rates of over 30\% compared to similarly trained single-vector retrievers using the same backbone, almost matching the performance of models with ten times more parameters~\citep{modernvlm}. 

Finally, while initially limited due to its considerable storage requirements, subsequent research has considerably improved the usability of late-interaction methods by targeting efficiency improvements, with  aggressive quantization in ColBERTv2~\citep{colbertv2}, better indexing methods such as PLAID~\citep{plaid} and WARP \citep{warp}, among others, vector count reduction via near-lossless pruning~\citep{tokenpooling,pruning} or using a fixed number of representative tokens\citep{constbert,metaembed}. The combination of these methods have led to multi-vector models being a viable option for many uses, and it currently stands as one of the main paradigms studied in neural IR research~\citep{colbertv2,colbertxm,colbertx,colpali,maskstudy,colberter,plaidrepro}.

\section{Theoretical Limitations of Current Methods}

ColBERT, and all existing related models building on it, such as multi-lingual~\citep{xlmcolbert,jacolbertv2.5} or multi-modal variants~\citep{colpali} use a simple mechanism to produce token-level representations: the hidden states of the final layer of a pre-trained backbone model, such as BERT~\citep{bert} or PaliGemma~\citep{paligemma} are passed through the most simple form a feedforward network can take: a single linear projection  $h(x) = xW$ where $W \in \mathbb{R}^{d \times k}$ to reduce their dimension from $d$ to $k$, with $k$ most most commonly set to 128~\citep{colbertv2,colpali,gtemoderncolbert}, before L2-normalizing the output of this projection:

\begin{align}
\hat{q}_i &= \frac{h(q_i)}{\|h(q_i)\|_2}, \quad 
\hat{d}_j = \frac{h(d_j)}{\|h(d_j)\|_2}
\end{align}

\subsection{MaxSim}
\label{sec:maxsim}

\subsubsection{Definition}

Using these embeddings, multi-vector retrieval techniques then compute relevance scores using the MaxSim operation\footnote{Or an approximation thereof~\citep{muvera,xtr}.}~\citep{colbert}. MaxSim is a simple operator where the cosine similarity between each query token and every document token is computed before discarding all similarities except the highest one for each query token (the \textit{max}imum \textit{sim}ilarity). Finally, all token-level maximum similarities are summed up, with this sum being used as the final relevance score assigned to a document for a given query. 

\begin{equation}
\begin{aligned}
\text{MaxSim}(q,d) &= \sum_{i=1}^m \max_{1 \leq j \leq n} \hat{q}_i^\top \hat{d}_j
\\[-4pt]
&\quad \text{where } m \text{ denotes the number of query tokens,}\\
&\quad \text{and } n \text{ denotes the number of document tokens.}
\end{aligned}
\label{eq:maxsim}
\end{equation}

\subsubsection{Maxsim's Gradient Flow}

Despite having empirically demonstrated strong performance, the MaxSim operator effectively creates a specific learning condition by limiting the information that flows back through the model. Indeed, let the winning document token for query token $i$ be:
\begin{equation}
j^*(i) = \argmax_{1 \leq j \leq n} \hat{q}_i^\top \hat{d}_j
\label{eq:argmax}
\end{equation}

Through the chain rule for max operations, we can observe that during the training phase of the model, gradients during backpropagation~\citep{backprop} will only flow through winning tokens:
\begin{equation}
\frac{\partial \text{score}}{\partial \hat{q}_i} = \hat{d}_{j^*(i)}, \quad
\frac{\partial \text{score}}{\partial \hat{d}_j} = 
\begin{cases}
\hat{q}_i & \text{if } j = j^*(i) \\
0 & \text{otherwise}
\end{cases}
\label{eq:sparse_grad}
\end{equation}

This creates an information bottleneck in the gradient flow. From Eq.~\eqref{eq:sparse_grad}, we can infer that the gradient with respect to each document token embedding~$\hat{d}_j$ is nonzero when and only when $j = j^*(i)$, which occurs when the token achieves the maximum similarity for at least one query token and is thus used by MaxSim. All other document tokens ($j \neq j^*(i)$) receive zero gradient, leading to them not contributing to learning during backpropagation. Similarly, each query token~$\hat{q}_i$ only receives gradient information from its corresponding winning document token~$\hat{d}_{j^*(i)}$. As a result, only a small subset of token pairs contribute to learning at each optimization step, effectively restricting the signal path for parameter updates. This selective flow of gradients through the ``winning'' pairs constitutes the key learning mechanism induced by the use of MaxSim during training.  For ease of referring to this concept while remaining readable, we refer to this effect as the ``winner-takes-all'' mechanism.

\subsection{Potential Limits of Token-level Linear Projections}

While computationally efficient, the single-layer projection used in existing multi-vector models applies the same transformation to every token, regardless of content or role in matching. Indeed, a linear head $h(x)=xW$ induces a single transformation matrix $W$ used uniformly for all tokens. After L2-normalization, cosine similarity is measured under a \emph{fixed} metric:
\begin{equation}
\label{eq:fixed_metric}
\mathrm{sim}(\hat q_i,\hat d_j)
\;=\;
\frac{(q_i W)(d_j W)^\top}{\|q_i W\|_2\,\|d_j W\|_2}
\;=\;
\frac{q_i^\top M\, d_j}{\sqrt{q_i^\top M q_i}\,\sqrt{d_j^\top M d_j}},
\qquad
M:=WW^\top \succeq 0.
\end{equation}

However, in practice, MaxSim rewards high peak similarities through its winner-takes-all mechanism, which can conflict with this mapping: Consider the trace constraint $\mathrm{tr}(M)=k$, which is enforced by dimensionality reduction and weight decay during training. With orthonormal directions $e_1, e_2, \ldots, e_d$ representing different semantic dimensions, we have:
\begin{equation}
\sum_{i=1}^d e_i^\top M e_i = \mathrm{tr}(M) = k
\end{equation}

For tokens aligned along direction $e_i$, their maximum achievable similarity after normalization is proportional to $e_i^\top M e_i$. To serve all token types adequately, $M$ must allocate some weight to every relevant direction, preventing it from concentrating strongly in any subset. This forced spreading theoretically yields lower peaks, as $M$'s eigenvalues are distributed rather than concentrated.

Under MaxSim's winner-takes-all supervision (Eq.~\eqref{eq:sparse_grad}), frequent winners pull $M$ toward their preferred directions, but the single $M$ must still maintain some support for all directions to avoid completely failing on certain token types. This creates a tension between the optimization objective which favours peaked distributions and the architectural constraint that which encourage spreading through a single global metric.

\section{Alternate ColBERT Projections and Their Expected Effects}
\label{sec:method}

The limitations highlighted above have not stopped ColBERT models and the associated MaxSim operators to empirically yield strong results across modalities, both in and out of domain. These results are in line with our assumption: even with single-matrix projections facing limitations, they are mitigated by the strong representation capabilities of the underlying Transformer-based~\citep{transformers} pre-trained backbone models, and the lack of a sharpening effect would not be sufficient to render performance non-competitive. 

However, we theorise that straightforward modifications, borrowed from the greater deep learning community practices, could help ColBERT models further alleviate the limitations of their naive projection mechanism. Specifically, we believe that factorization benefits that arise from modest model depth alone would yield considerable cross-domain improvements.

We further propose the use of residual skip-connections as part of these multi-layered projection, to allow the projection to focus on producing a sharpening effect while being able to rely on the backbone models' original representations to stabilise the final embeddings.

We also investigate the use of various forms of non-linearities, through the use of common non-linear activation functions, as well Gated Linear Units~\citep{glu}, a widely-used alternative to the traditional feedforward block that introduces an additional non-linear gating~\citep{glushazeer}. 

All of these mechanisms are commonly used as part of modern deep model architectures, with model depth thought to contribute to downstream performance more than model width\footnote{At the cost of efficiency tradeoffs at very high layer counts.}~\citep{widevsdeep,widevsdeep2} and various forms of non-linearity being considered key to model feedforward layers~\citep{glu,relu,gelu,silu}.

\subsection{Depth Introduces Sharpening-Improving Factorization}
\label{sec:ffn}

Multi-layer feedforward networks (FFNs) are constructed by simply stacking linear layers, with an activation function applied to the output of a layer before being passed to the next one. In their simplest form, the activation function can be a simple Identity function, in which case the output of an intermediate layer in dimension $m$ is passed as-is to the next:
\begin{equation}
h_{\text{FFN}}(x) = \phi(xW_1)W_2, \quad W_1 \in \mathbb{R}^{d \times m}, W_2 \in \mathbb{R}^{m \times k} \quad \text{where } \phi \text{ is an activation function}
\end{equation}

It is also common for multilayered feedforward blocks to adopt a so-called bottleneck design, following the original Transformer~\citep{transformers}, where the first projection expands to a higher dimension before the final layer projects back to the desired output dimension. We define the projection scale $\rho$ (rho), which controls the intermediate dimension. Given an input $\mathbf{x} \in \mathbb{R}^{d}$, the bottleneck operations are:
\begin{align}
    h &= \phi(W_\text{up} x + b_\text{up}), \quad W_\text{up} \in \mathbb{R}^{d \times m} \\
    y &= W_\text{down} h + b_\text{down}, \quad W_\text{down} \in \mathbb{R}^{d \times m}
\end{align}
where $m$ is an intermediate dimension controlled by $\rho$ and defined as $\rho \times d$, $\mathbf{h} \in \mathbb{R}^{dm}$ is the intermediate representation with expanded dimensionality. The first projection (upcasting) expands the dimension from $d$ to $m$, while the second projection (downcasting) reduces it back to $d$ in the case of an intermediate layer, or $k$ if it is the final layer of a down-projection FFN, as it is in the ColBERT context.

Even with the activation function $\phi$ defined as the identity function rather than a non-linear activation function, we suggest that the factorization introduced by the addition of an additional layer to the ColBERT projection head would lead to two improvements which benefit MaxSim: increased spectral concentration and better gradient aggregation.

\subsubsection{Spectral concentration} All standard training methods for ColBERT employ weight decay~\citep{wd}, which applies $L_2$ regularization to model weights. In practice, weight decay on factors $\|W_1\|_F^2 + \|W_2\|_F^2$ implicitly regularizes the nuclear norm $\|W_1W_2\|_*$, encouraging low effective rank. This occurs because for any factorization:
\begin{equation}
\|W_1W_2\|_* \leq \|W_1\|_F \|W_2\|_F \leq \frac{1}{2}(\|W_1\|_F^2 + \|W_2\|_F^2)
\end{equation}

For a rank-$r$ approximation with singular values $\sigma_1 \geq \ldots \geq \sigma_r$, the trace constraint from dimensionality reduction gives $\sum_i \sigma_i^2 = k$. Lower rank solutions concentrate this ``budget'' into fewer, larger singular values, yielding:
\begin{equation}
    \max_{\|v\|=\|u\|=1} v^T W_1W_2 u = \sigma_1 \gg \frac{k}{\sqrt{d}}
\end{equation}

This concentration effect encourages projections to be concentrated towards fewer singular directions, leading to sharper, or ``peakier'', token embeddings with higher potential maximum similarities, thus directly benefiting from MaxSim's winner-takes-all effect.

\subsubsection{Better handling of gradient aggregation} The factorized structure improves conditioning for aggregating the many sparse rank-1 updates from MaxSim. Under MaxSim's gradient flow (Eq.~\eqref{eq:sparse_grad}), each winning pair $(q_i, d_{j^*(i)})$ contributes a rank-1 update to the projection. With a single matrix $W$, these updates directly compete:
\begin{equation}
\Delta W \propto \sum_{\text{winners}} d_{j^*(i)} q_i^T
\end{equation}
In contrast, factorization $W_1W_2$ creates an intermediate representation space of dimension $h$ where updates are first aggregated in $W_1$ before being projected by $W_2$. This two-stage process allows the model to learn shared intermediate features that benefit multiple token types, rather than forcing each token type to claim dimensions in the final space. The intermediate bottleneck could act as a regularizer, encouraging the discovery of composable features that can be combined differently for different semantic types, without damaging the sharpening that is beneficial for MaxSim.

\subsection{Residual connections}
\label{sec:skipco}

Residual connections~\citep{skipconnection} are frequently used in deep learning, as they have been demonstrated to improve training stability and downstream performance. A residual connection effectively adds the input to the projection's output, with a learned multiplier $\alpha$:
\begin{equation}
h_{\text{residual}}(x) = x + \alpha \cdot g(x), \quad \text{where } g(x) = xW_1W_2
\end{equation}

In the context of multi-vector retrieval, we believe that residual connections could potentially offer the benefit of enabling greater role decomposition in the learned projections. The effective metric induced by this formulation becomes Eq.~\eqref{eq:res}, where $W = W_1W_2$ for notational simplicity:
\begin{equation}
\label{eq:res}
h_{\text{residual}} = (I + \alpha W)(I + \alpha W)^T = I + \alpha(W + W^T) + \alpha^2 WW^T
\end{equation}
This decomposition highlights two complementary components: the identity $I$ preserves the semantic geometry of the fine-tuned backbone model's until the final projection, while the learned term $\alpha W$ theoretically gains greater freedom to focus on amplifying distinctive tokens during the training process by creating an interaction between the original and learned representations. In the context of MaxSim, this allows the model to selectively boost winners through the learned components. We theorise that this can potentially lead to higher peak similarities without sacrificing performance on non-dominant token types.

\subsubsection{Residual Connection In 2-Layer FFNs}

We make a note that when implementing residual connections with a 2-layer feedforward projection, which effectively projects the input dimension $d$ to intermediate dimension $m$, then immediately down to output dimension $k$, we adopt a residual connection inspired by ResNets~\citep{skipconnection} to ensure that the individual dimensions match. In effect, this means that we upcast the input using an additional upcasting layer, whose weights are initialized as an identity matrix to modify the input as little as possible while performing the dimension mapping. We do so as it would otherwise be impossible for us to evaluate the potential benefit of residual connections at a depth of 2 projection layers, as we would require an additional intermediate downcasting back to $d$ to be able to create a residual connection with the input of dimension $m$.

\subsection{Non-Linearity and Gating}
\label{sec:nonlin}

Introducing non-linearity into feedforward layers is a common practice when designing deep model architectures. In our context, we believe it could potentially enable \emph{input-dependent} transformations that can selectively emphasize token dimensions. Non-linearity can be injected either via activation functions applied to the output of intermediate layers, with widely used functions in NLP such as ReLU~\citep{relu}, SiLU~\citep{silu}, GELU~\citep{gelu}), or via  gated blocks such as Gated Linear Units (GLU)~\citep{glu}.

\subsubsection{Multi-layer block with activations.}
In Section~\ref{sec:ffn}, we introduced the use of multi-layer feedforward networks (FFN) with the use of the identity activation function, where no modifier is applied to model outputs. An activation function can be introduced into this block to introduce non-linearity:
\begin{equation}
\label{eq:mlp_proj}
h_{\text{FFN}}(x) \;=\; \phi(xW_1)\,W_2, 
\qquad
W_1 \in \mathbb{R}^{d \times h},\;\; W_2 \in \mathbb{R}^{h \times k},
\end{equation}
followed by L2-normalization as in Eq.~(1). Particularly relevant to multi-vector retrieval, the use of a non-linear activation $\phi$ induces an input-dependent Jacobian $J$, which means that different tokens are affected differently by the operation:
\begin{equation}
\label{eq:jacobian}
J_{h_{\text{FFN}}}(x) 
\;=\; \frac{\partial h_{\text{FFN}}(x)}{\partial x}
\;=\; W_1\, \mathrm{Diag}\!\big(\phi'(xW_1)\big)\, W_2.
\end{equation}
Equation~\eqref{eq:jacobian} shows that $\phi'(xW_1)$ gates columns of $W_1$ before mixing by $W_2$. After normalization, this leads the local cosine geometry to depend on $J_{h_{\text{FFN}}}(x)$ through the $J^\top J$ it induces. Theoretically, this could enable token-specific emphasis and result in greater similarity peaks, which would subsequently be rewarded by MaxSim.

\subsubsection{Gated Linear Units (GLU)}
\label{sec:glu}
GLUs introduce a multiplicative gate that modulates a value stream:
\begin{equation}
\label{eq:glu_general}
h_{\text{GLU}}(x) \;=\; \big(xW_v\big) \;\odot\; \psi\!\big(xW_g\big),
\qquad
W_v \in \mathbb{R}^{d \times k},\;\; W_g \in \mathbb{R}^{d \times k},
\end{equation}
where $\psi$ is the gating nonlinearity and $\odot$ denotes elementwise multiplication. Originally, GLU layers were introduced with the use of a sigmoid gated, expressed as $\psi=\sigma$ (sigmoid)~\citep{glu}. 

Subsequent work has shown benefits from alternative gates which replace the sigmoid with common activation functions, creating variants such as ReGLU ($\psi=\mathrm{ReLU}$), GEGLU ($\psi=\mathrm{GELU}$), and SwiGLU ($\psi=\mathrm{SiLU}$)~\citep{glushazeer}. However, performance between different GLU variants have been shown to vary, and the reasons for such variations are currently poorly understood.

\paragraph{Non-linearity even with identity gate.}
Finally, it is worth noting that even with an identity gate ($\psi(u)=u$), GLU introduces non-linearity nonetheless, where GLU with identity gating reduces to a \emph{bilinear} layer that introduces pairwise feature interactions $x_i x_j$:
\begin{equation}
\label{eq:glu_identity}
\big(h_{\text{GLU}}(x)\big)_k 
\;=\; \big(x^\top W_v^{(:,k)}\big)\,\big(x^\top W_g^{(:,k)}\big)
\;=\; x^\top \!\Big(W_v^{(:,k)} {W_g^{(:,k)}}^{\!\top}\Big)\, x,
\end{equation}

Therefore, in our context, even an identity-gated GLU would result in introducing non-linearity through quadratic feature interactions. This mechanism could potentially capture more complex semantic relationships between token dimensions than linear projections alone, creating a situation in which the projection could learn that certain feature combinations are particularly indicative of relevance, should that be the case in the backbone model's hidden states. Under the lense of MaxSim, this means the projection could learn to amplify similarities when specific pairs of features co-occur, potentially creating sharper peaks.

\subsubsection{Potential Effects of Non-Linearity}

We theorize that the interactions of Non-Linearity with the MaxSim operator and the gradient flow constraints it introduces, as presented in Section~\ref{sec:maxsim}, are both positive and negative in regards to introducing downstream performance.

\paragraph{Potentially Increased Sharpening}  Eqs.~\eqref{eq:jacobian} and \eqref{eq:glu_general} show that both non-linearities and gating enable input-dependent reweighting, which has the potential to concentrate each token’s mass along a few decisive directions. Theoretically, this could increase dynamic range of $\hat q_i^\top \hat d_j$ and producing clearer winners $j^*(i)$, facilitating the learning process. Additionally, because we use L2-normalized, absolute scale changes are suppressed, but \emph{directional} changes induced by non-linear activations or GLU still alter cosine similarity. Locally, the effective metric is governed by $J^\top J$ (for non-gated FFNs) or the product-rule Jacobian of Eq.~\eqref{eq:glu_general} for GLU, facilitating sharpening via directional emphasis, rather than purely scalar.

\paragraph{Potential Negative Effects} The harder sparsity introduced by non-linearity and gating, on the other hand, can result in an over-sharpening which would increase winner instability, risking amplifying the winner-takes-all bottleneck of Eq.~\eqref{eq:sparse_grad}, which, due to how backpropagation works, would decrease the odds of the model training successfully converging~\citep{backprop}. Additionally, certain types of non-linearity can dampen the learning signal: for example, sigmoid gates risk saturation in extreme input regions, while ReLU can zero gradients for negative inputs. These mechanisms could block the signal from reaching earlier layers, even when a pair is selected by MaxSim, thus hindering the learning process.

\subsection{Validating Theoretical Learning Properties Requires Empirical Evidence} 

In this section, we have presented multiple mechanisms and proposed theoretical justifications for their impact on MaxSim performance. Some of these mechanisms, as is the case for non-linearity, have conflicting properties for this setting, which is common in deep learning. 

Even for non-conflicting properties, the learning mechanisms induced by backpropagation ultimately remain largely a black box. As such, while theories can be made, it is difficult to predict which learning effect will have the greater impact during model training. Famously, the paper introducing the use of GLU variants end on this often-quoted note:

\begin{quote}
\emph{``We offer no explanation as to why these architectures seem to work; we attribute their success, as all else, to divine benevolence.''}\citep{glushazeer}
\end{quote}

While tongue-in-cheek, this remark highlights that, within deep learning, the effect of theoretically sound, if potentially conflicting, modifications are frequently only understood through the empirical lense. Indeed, deep learning remains largely a black box~\citep{blackbox}, with empirical performance serving as the most common validator. This is especially applicable in situations such as ours, where some proposed modifications have both positive and negative theoretical properties.

As such, in our experiments, we introduce all modified projection layers presented above to our models, including all forms of non-linearity, in order to demonstrate their empirical effect, whether beneficial or harmful.

\section{Experimental Setting}

Our aim is to be thorough in our experiments, evaluating all the combinations of settings described in Section~\ref{sec:method} in a way that is both significant and applicable to state-of-the-art training methods, to demonstrate that potential improvements are not dependent on a weak baseline. In this section, we present the training decisions made to ensure both of these while keeping compute requirements reasonable.

\subsection{Implementation}

Building on the justification presented in Section~\ref{sec:method}, we present an experimental framework which will allow us to measure the effects of our proposed model modifications. Specifically, we seek to measure the impact of introducing various projection blocks to replace the currently used linear projection.

We extend the PyLate library~\citep{pylate}, a widely-used framework for the training and evaluation of multi-vector retrieval models, to support modular projection blocks. Specifically, we implement the ability to control the following parameters:
\begin{itemize}
    \item \textbf{Projection depth}: How many feedforward blocks should be used for the projection.
    \item \textbf{Gated Linear Units}: Whether to use GLU layers instead of traditional feedforward layers without gating.
    \item \textbf{Residual connection}: Whether there should be a skip-connection between layers or not, as presented in Section~\ref{sec:skipco}.
    \item \textbf{Activation function}: The activation function to be applied to the output of non-final layers.
    \item \textbf{Projection Scale}: As presented in Section~\ref{sec:ffn}, it is common for feedforward layer to adopt a larger scale for intermediate projections. For the sake of thoroughness, we ablate the effect of a using non-scaled up projections, as well as projections where the intermediate layer's dimension is twice that of the input dimension, on retrieval performance.

\end{itemize}

Despite many studies exploring activation functions, the empirical performance of different activation functions remains fluctuating and largely task-dependent, without, as of yet, clear general patterns or theoretical reasons as to why performance fluctuates~\citep{activationfunctions}. Identifying these exact factors remains out of the scope of this study, and we follow existing practice~\citep{glushazeer} in comparing empirically performance among multiple activations, among the most commonly used ones for natural language processing tasks: Identity (no activation), ReLU, GELU, SiLU, and, for the sake of thoroughness, GLU layers with their original sigmoid gating, all briefly presented in Section~\ref{sec:nonlin}.

\subsection{Training Setting}

\paragraph{Base Model} Smaller models, such as MiniLM~\citep{minilm}, have repeatedly been demonstrated to be well-suited for retrieval tasks, reaching strong performance. This is especially true for ColBERT models, where recent empirical results have shown that even a 4-million parameter ColBERT model could be competitive with models over 30 times larger~\citep{4mcolbert}. Moreover, it is common to conduct experiments on smaller models, with scaling laws showing that their results are extremely strongly correlated with the results of larger variants~\citep{scalinglaws}. As such, we choose to use the 32M parameter variant of Ettin as our backbone model. Ettin~\citep{ettin} is an improved reproduction across model sizes of ModernBERT~\citep{modernbert}, itself a variant of the original BERT~\citep{bert} incorporating recent advances in model training. 

\paragraph{Training Setting} We conducted limited sweeps over hyperparameters on a handful of settings. Our findings largely match previous research, with a batch size of 64, a learning rate of $1e-4$ with a linear decay schedule following a warmup phase for 10\% of total training steps reaching consistently strong performance. As such, we adopt these settings for all experiments. Following ColBERT training efforts since ColBERTV2~\citep{colbertv2}, we adopt a knowledge distillation loss where the training objective is to minimize the difference between the score distribution of the student and teacher models. We follow standard existing practice use Kullback–Leibler divergence (KL-Div) between the student and teacher scores as our loss function, which has empirically been shown to be well suited for ColBERT training~\citep{jacolbertv2.5,colbertv2} and retrieval models in general~\citep{kldiv1,kldiv2}.

\paragraph{Data} To increase applicability of our method to the real-world, we train our model using a 640,000 sample of the data commonly used to train current state-of-the-art ColBERT models~\citep{answeraicolbert,4mcolbert,gtemoderncolbert}. This training set is effectively a downsample of ColBERTV2's large original corpora of 64-way training tuples from MS Marco~\citep{msmarco}, each composed of a query, a positive example, and 63 negative examples mined via MiniLMv2~\citep{colbertv2,minilm} with teacher scores generated by a reranker. Our sampled set instead uses 640,000 randomly selected 16-way tuples, reducing the number of negatives to 15, with scores generated by bge-reranker-m3~\citep{bgem3}. This downsampling has previously been shown to be sufficient to yield results that are significantly correlated with performances obtained when training on 10x more data~\citep{jacolbertv2.5,answeraicolbert}, while significantly lowering training compute requirements.

\subsection{Consistency}

The reproducibility of experiments in machine learning is an often-discussed topic, with studies showing that reported results are often, even if involuntarily, cherry-picked, with more neutral evaluation methods showing different results~\citep{showyourwork}. 

Particularly, it has been demonstrated that robustness-across-random-seeds is an important component of demonstrating the suitability of new methods, with sharp single-seed improvements showing a greatly diminished effect with multi-seed comparisons~\citep{reproinml,randomseeds}. This effect is observed across all of deep learning, with computer vision models showing significant downstream performance variance across training runs where random seeding is the only changed parameter~\citep{jordanvariance}.

In retrieval, it has been shown that random seeds can greatly impact the downstream performance of QA answer retrieval tasks, with relative performance variations of over 10\% being observed across seeds, potentially negatively altering the course of future research as a poorly chosen seed could be sufficient to take a method from state-of-the-art performance to noticeably trailing existing methods~\citep{qarandomseed}.

While the computational requirements of machine learning training mean large-sample size significance studies are difficult, the authors highlight the importance of taking reasonable steps to ensure more reproducible comparisons, such as evaluating methods across multiple seeds. 

As such, we conduct all training and evaluation runs five times, using five individual random seeds for PyTorch seeding, parameter initialization and dataset shuffling: $1$, $42$, $1337$, $1789$ and $1861$. All results are reported as the mean of the checkpoints resulting from the five seeds, across three separate indexing runs each to eliminate indexing variance.

\subsection{Evaluation Settings}

\paragraph{Data} We report results across a set of commonly used, standardised benchmarks: TREC-DL19 and TREC-DL20, as well as the high-quality search subsets of the BEIR evaluation suite~\citep{beir}: SciFact~\citep{scifact}, TREC-Covid~\citep{treccovid}, FiQA2018~\citep{fiqa} and NFCorpus~\citep{nfcorpus}. We select these benchmarks as they cover multiple domains and are widely used and generally considered to be high quality collections, without incurring the computational cost of running full BEIR evaluations across multiple seeds for all evaluated settings.

\paragraph{Indexing and Searching} All evaluations are ran using the standardised ColBERTv2~\citep{colbertv2}+PLAID~\citep{plaid} indexing method. PLAID is an optimized index type built upon an inverted file index coupled with aggressive product quantization, allowing for fast multi-vector retrieval while reducing index sizes. We employ 4-bit quantization for individual token vectors follow the optimal parameters identified by a recent thorough PLAID reproduction study~\citep{plaidrepro} at inference time. Query length is set to 32 and document length to 300, following commonly used settings~\citep{colbert}. All indexes are created and searched through using the PyLate library~\citep{pylate}, with DL-19 and DL-20 loaded separately via ir-datasets~\citep{irdatasets}.

\section{Experimental Results}
\label{sec:results}

In this section, we will empirically explore the performance of our various proposed projection modification. 

We will first highlight a high-level overview of so-called ``canonical'', that is, using widely used default parameters, FFN and GLU blocks, comparing their performance to that of the commonly used single-layer linear projection baseline.

Subsequently, we will present the results of targeted evaluations seeking to further explore individual factors that impact well-performing model variants, such as the choice of activation function (Sec.~\ref{sec:res_act}), the use of residual connections(Sec.~\ref{sec:res_skipco}) and of a higher intermediate projection dimension(Sec.~\ref{sec:res_upscale}).

\subsection{Overall Results}

\begin{table}
\centering
\caption{Main results showing a comparison of the linear baseline with various depth for the most common settings for each FFN family across model depths. All results reported are NDCG@10 averaged across 5 training runs. Results in \textbf{bold} are the best overall results and results \underline{underlined} are results which outperform the baseline projection. \textdagger denotes statistical significance with $p<0.05$, with more information on significance provided in ~Sec\ref{sec:significance}.}
\label{tab:main_results}
\begin{tabular}{lccccccc}
\toprule
Model & DL19 & DL20 & COVID & SciFact & NFC & FiQA & Avg. \\
\midrule
\textbf{Baseline} &  &  &  &  &  &  &  \\
Linear Projection & 0.6857 & 0.7081 & 0.6831 & 0.6728 & 0.3355 & 0.3311 & 0.5694 \\
\midrule
\textbf{FFN (identity)} &  &  &  &  &  &  &  \\
Depth 2 & \underline{\textbf{0.7095}}\textsuperscript{\textdagger} & 0.7040 & \underline{0.7402}\textsuperscript{\textdagger} & \underline{\textbf{0.7076}}\textsuperscript{\textdagger} & \underline{0.3382} & \underline{\textbf{0.3455}}\textsuperscript{\textdagger} & \underline{\textbf{0.5908}}\textsuperscript{\textdagger} \\
Depth 3 & \underline{0.6931}\textsuperscript{\textdagger} & 0.7012 & \underline{0.7452}\textsuperscript{\textdagger} & \underline{0.7044}\textsuperscript{\textdagger} & \underline{0.3369} & \underline{0.3369}\textsuperscript{\textdagger} & \underline{0.5861}\textsuperscript{\textdagger} \\
Depth 4 & \underline{0.6911}\textsuperscript{\textdagger} & \underline{0.7082} & \underline{0.7338}\textsuperscript{\textdagger} & \underline{0.6891}\textsuperscript{\textdagger} & \underline{0.3356} & \underline{0.3414}\textsuperscript{\textdagger} & \underline{0.5864}\textsuperscript{\textdagger} \\
\midrule
\textbf{GLU (sigmoid)} &  &  &  &  &  &  &  \\
Depth 2 & \underline{0.6944}\textsuperscript{\textdagger} & 0.7022 & \underline{0.7448}\textsuperscript{\textdagger} & \underline{0.7049}\textsuperscript{\textdagger} & \underline{0.3383} & \underline{0.3435}\textsuperscript{\textdagger} & \underline{0.5880}\textsuperscript{\textdagger} \\
Depth 3 & \underline{0.6995}\textsuperscript{\textdagger} & \underline{\textbf{0.7108}} & \underline{0.7274} & \underline{0.7002}\textsuperscript{\textdagger} & \underline{0.3388} & \underline{0.3406}\textsuperscript{\textdagger} & \underline{0.5862}\textsuperscript{\textdagger} \\
Depth 4 & \underline{0.6966} & 0.7061 & \underline{\textbf{0.7465}}\textsuperscript{\textdagger} & \underline{0.7010}\textsuperscript{\textdagger} & \underline{\textbf{0.3399}} & \underline{0.3461}\textsuperscript{\textdagger} & \underline{0.5893}\textsuperscript{\textdagger} \\
\bottomrule
\end{tabular}
\end{table}

Table~\ref{tab:main_results} presents a comparison of the performance of the commonly used linear projection against a set of varying depths FFN and GLU projections. For the ease or readability, we provide only the most standardised version of these projection blocks: residual connections are used, a projection scale of 2.0, i.e. twice the input dimension, is used in intermediate layers and we do not use a non-linear activation function for the FFN blocks, while we use the canonical sigmoid gate for GLU blocks.

The results starkly demonstrate that these projection variants significantly outperform the baseline projection on all datasets evaluated, with the exception of DL20 where the performance of some projections is very slightly inferior to that of the baseline. In this context, it is worth noting that both DL19 and DL20 are in-domain datasets, using the same MS Marco~\citep{msmarco} document collection that was used to train the model, while the other four datasets are fully out-of-domain. Under this light, we can note that alternate projections observe no degradation, and even gains on DL19, while in-domain, while noticeably improving performance on 3 out of 4 out-of-domain evluations and reaching moderate gains on the fourth.

The significant gains achieved on Trec-COVID, SciFAct and FiQA appear to support the theory expressed in Section~\ref{sec:ffn}, in which we propose that alternate projections would be particularly useful in facilitating the representation of domain-specific vocabulary, thus increasing performance.

Overall, we note that these results support the idea that the use of alternate projection is an underexplored, ``almost-free lunch'' to improve the retrieval performance of ColBERT models.

\subsubsection{Significance of the Results}
\label{sec:significance}

It is hard to assess true statistical significance of model variations without incurring significant training costs, as sample sizes remain very modest. While many of our results above appear statistically significant, the low number of observed points create high variance. To further highlight the effect of our variants, we ran further training runs with selected well-performing variants, FFN at depth 2 and GLU at depth 4, as well as the baseline projection, on five additional random seeds. We then evaluated these new checkpoints to gather additional information on significance. We present  the p-values resulting from paired two-sided t-tests results in Table~\ref{tab:significant}.

\begin{table}[h!]
  \centering
  \caption{Paired two-sided $t$-test $p$-values comparing model variations w.r.t. the linear baseline. Bold entries indicate $p < 0.05$.}
  \label{tab:significant}
  \begin{tabular}{lcccccc}
  \toprule
  Model & DL19 & DL20 & COVID & SciFact & NFC & FiQA \\
  \midrule
  FFN Depth 2 & $\mathbf{0.005}$ & $0.311$ & $\mathbf{0.049}$ & $\mathbf{0.009}$ & $0.488$ & $\mathbf{0.015}$ \\
  \midrule
  GLU Depth 4 & $\mathbf{0.037}$ & $0.115$ & $\mathbf{0.040}$ & $\mathbf{0.009}$ & $0.281$ & $\mathbf{0.021}$ \\
  \bottomrule
  \end{tabular}
  \end{table}

This analysis highlights two factors: performance variations on NFcorpus are not significant, due to the very small variations across models. We hypothesize that demonstrating statistical significance on on NFCorpus would require an extremely large number of runs, as even the performance of state-of-the-art models on NFCorpus on the MTEB leaderboard~\citep{mteb} shows that even large swing in overall model performance result in only modest increases on this dataset. Secondly, performance on DL20, a dataset where the linear projection outperformed our alternate projection in our original reports, is statistically insignificant, with large variations across training runs. Apart from these two datasets, we observe that performance variations on all four other datasets are statistically significant. Interestingly, DL19 performance improvements for GLU at depth 4, which fell short of the significance threshold previously, become significant when accounting for these additional data points.

\subsection{Activation Functions and Non-Linearity}
\label{sec:res_act}

\begin{table}
\centering
\caption{Comparison of activation functions for FFN and GLU projection variants with all other parameters kept equal, averaged across five model checkpoints. All results are NDCG@10. Results in \textbf{bold} are the best overall per column, and results \underline{underlined} indicate they outperform the baseline.}
\label{tab:act_ffn_glu}
\begin{tabular}{lccccccc}
\toprule
Model & DL19 & DL20 & COVID & SciFact & NFC & FiQA & Avg. \\
\midrule
\textbf{Baseline} &  &  &  &  &  &  &  \\
Linear Projection & 0.6857 & 0.7081 & 0.6831 & 0.6728 & 0.3355 & 0.3311 & 0.5694 \\
\midrule
\textbf{FFN} &  &  &  &  &  &  &  \\
$FFN_{Identity}$ & \underline{\textbf{0.7095}} & 0.7040 & \underline{\textbf{0.7402}} & \underline{\textbf{0.7076}} & \underline{\textbf{0.3382}} & \underline{\textbf{0.3455}} & \underline{\textbf{0.5908}} \\
$FFN_{ReLU}$ & \underline{0.6929} & 0.7005 & \underline{0.7237} & \underline{0.7030} & \underline{0.3381} & \underline{0.3436} & \underline{0.5836} \\
$FFN_{GELU}$ & \underline{0.6957} & 0.6967 & \underline{0.6905} & \underline{0.7048} & \underline{0.3357} & \underline{0.3437} & \underline{0.5778} \\
$FFN_{SiLU}$ & 0.6791 & 0.6968 & \underline{0.7124} & \underline{0.7009} & 0.3354 & \underline{0.3326} & \underline{0.5762} \\
\midrule
\textbf{GLU} &  &  &  &  &  &  &  \\
$GLU_{Sigmoid}$ & \underline{0.6944} & 0.7022 & \underline{\textbf{0.7448}} & \underline{0.7049} & \underline{0.3383} & \underline{0.3435} & \underline{0.5880} \\
$GLU_{Identity}$ & \underline{0.6898} & \underline{\textbf{0.7105}} & \underline{0.7394} & \underline{0.6993} & \underline{0.3381} & \underline{0.3431} & \underline{0.5869} \\
$GLU_{ReLU}$ & \underline{0.6891} & 0.7034 & \underline{0.7196} & \underline{0.6999} & \underline{0.3377} & \underline{0.3428} & \underline{0.5829} \\
$GLU_{GELU}$ & \underline{\textbf{0.7082}} & 0.7038 & \underline{0.7371} & \underline{\textbf{0.7061}} & 0.3380 & \underline{\textbf{0.3449}} & \underline{0.5892} \\
$GLU_{SiLU}$ & \underline{0.6969} & 0.7007 & \underline{0.7306} & \underline{0.7008} & \underline{0.3372} & \underline{0.3409} & \underline{0.5845} \\
\bottomrule
\end{tabular}
\end{table}

Table~\ref{tab:act_ffn_glu} presents the results of varying activation functions, with all other parameters being fixed to the best performing depth 2 variants presented in Table~\ref{tab:main_results}.

For FFN blocks, the use of activation function appears to be a net negative in terms of performance, across all datasets. While all activation functions continue to outperform the baseline, the gains are less pronounced. As such, it seems to indicate that the potentially sharpening effects of adding non-linearity do not outweigh their potential negative effects and rather ends up dampening the positive effects of other modifications.

For GLU blocks, which as indicated in Section~\ref{sec:glu} are non-linear no matter the activation function, it seems that the choice of activation function has only moderate impact, with all variants reaching broadly similar results, even if GELU pulls slightly ahead. Interestingly, GLU variants, while ultimately all outperformed by the $FFN_{identity}$ variant, reach more consistent results than non-linear FFN blocks and outperform the baseline in all evaluated settings. This seems to suggest that GLU layers do improve the quality of representations, although in a way directly tied to the gating mechanism.

Overall, these results appear to indicate that non-linear activation functions do not, overall, contribute to improving the projection quality of multi-vector retrieval models.

\subsection{Upscaling}
\label{sec:res_upscale}

\begin{table}
\centering
\caption{Comparison of projection scale $\rho$ across depths for FFN and GLU projection variants with all other parameters kept equal. All results are NDCG@10. Results in \textbf{bold} are the best overall per column, and results \underline{underlined} outperform the baseline.}
\label{tab:depth_scale}
\begin{tabular}{lccccccc}
\toprule
Model & DL19 & DL20 & COVID & SciFact & NFC & FiQA & Avg. \\
\midrule
\textbf{Baseline} &  &  &  &  &  &  &  \\
Linear Projection & 0.6857 & 0.7081 & 0.6831 & 0.6728 & 0.3355 & 0.3311 & 0.5694 \\
\midrule
\textbf{FFN (identity)} &  &  &  &  &  &  &  \\
$\rho{=}1$, Depth 2 & \underline{0.6905} & 0.7003 & \underline{0.6881} & \underline{0.7034} & \underline{0.3391} & \underline{0.3391} & \underline{0.5767} \\
$\rho{=}2$, Depth 2 & \underline{\textbf{0.7095}} & 0.7040 & \underline{0.7402} & \underline{0.7076} & \underline{\textbf{0.3382}} & \underline{\textbf{0.3455}} & \underline{\textbf{0.5908}} \\
$\rho{=}1$, Depth 3 & \underline{0.6859} & 0.6921 & 0.6488 & \underline{0.6994} & 0.3351 & 0.3322 & 0.5656 \\
$\rho{=}2$, Depth 3 & \underline{0.6931} & 0.6997 & \underline{\textbf{0.7452}} & \underline{0.7044} & \underline{0.3369} & \underline{0.3369} & \underline{0.5861} \\
$\rho{=}1$, Depth 4 & 0.6121 & 0.6037 & 0.6325 & \underline{0.6891} & 0.3281 & 0.3186 & 0.5307 \\
$\rho{=}2$, Depth 4 & \underline{0.6911} & \underline{\textbf{0.7082}} & \underline{\textbf{0.7338}} & \underline{\textbf{0.7081}} & \underline{0.3356} & \underline{\textbf{0.3414}} & \underline{0.5864} \\
\midrule
\textbf{GLU (GELU)} &  &  &  &  &  &  &  \\
$\rho{=}1$, Depth 2 & \underline{0.6940} & \underline{\textbf{0.7111}} & \underline{\textbf{0.7405}} & \underline{\textbf{0.7056}} & \underline{\textbf{0.3366}} & \underline{0.3430} & \underline{0.5885} \\
$\rho{=}2$, Depth 2 & \underline{0.6979} & 0.6987 & \underline{0.7151} & \underline{0.6990} & \underline{\textbf{0.3373}} & \underline{0.3401} & \underline{0.5813} \\
$\rho{=}1$, Depth 3  & 0.6642 & 0.6400 & 0.6601 & \underline{0.6867} & 0.3313 & 0.3206 & 0.5505 \\
$\rho{=}2$, Depth 3 & \underline{\textbf{0.7042}} & 0.7025 & \underline{0.7371} & \underline{\textbf{0.7056}} & \underline{\textbf{0.3392}} & \underline{\textbf{0.3448}} & \underline{0.5889} \\
$\rho{=}1$, Depth 4 & \underline{0.6923} & 0.7000 & \underline{0.6903} & \underline{0.6983} & 0.3335 & \underline{0.3315} & \underline{0.5743} \\
$\rho{=}2$, Depth 4 & \underline{0.7005} & \underline{\textbf{0.7099}} & \underline{0.7197} & \underline{\textbf{0.7099}} & \underline{0.3357} & 0.3308 & \underline{0.5844} \\
\bottomrule
\end{tabular}
\end{table}

\begin{table}[!ht]
\centering
\caption{Comparison of models with and without residual connections across projection scales $\rho$ for select FFN and GLU configurations. $\Delta$Avg denotes the performance difference between settings, with all else being kept equal (Residual $-$ No Residual). Values outperforming the baseline are \underline{underlined}, and the overall best result is in \textbf{bold}.}
\label{tab:residual_comparison_full}
\begin{tabular}{lcccccccc}
\toprule
Model & DL19 & DL20 & COVID & SciFact & NFC & FiQA & Avg. & $\Delta$ \\
\midrule
\textbf{Baseline} &  &  &  &  &  &  &  \\
Linear Projection & 0.6857 & 0.7081 & 0.6831 & 0.6728 & 0.3355 & 0.3311 & 0.5694 & -- \\
\midrule
\textbf{FFN} &  &  &  &  &  &  &  &  \\
\textbf{Depth 2} &  &  &  &  &  &  &  &  \\
$\rho{=}1.0$ No Residual & \underline{0.6907} & 0.7001 & \underline{0.6883} & \underline{0.7032} & \underline{0.3390} & \underline{0.3390} & \underline{0.5769} & -- \\
$\rho{=}1.0$ Residual & \underline{0.6905} & 0.7003 & \underline{0.6881} & \underline{0.7034} & \underline{0.3391} & \underline{0.3391} & \underline{0.5767} & $-$0.0002 \\
\addlinespace
$\rho{=}2.0$ No Residual & 0.6875 & 0.7030 & \underline{0.7336} & \underline{0.6954} & \underline{0.3368} & \underline{0.3342} & \underline{0.5707} & -- \\
$\rho{=}2.0$ Residual & \underline{0.7095} & \underline{0.7040} & \underline{0.7402} & \underline{0.7076} & \underline{0.3382} & \underline{0.3455} & \textbf{\underline{0.5908}} & $+$0.0201 \\
\addlinespace
\textbf{Depth 3} &  &  &  &  &  &  &  &  \\
$\rho{=}1.0$ No Residual & 0.6871 & 0.6929 & 0.6493 & \underline{0.6978} & \underline{0.3349} & 0.3315 & 0.5889 & -- \\
$\rho{=}1.0$ Residual & 0.6859 & 0.6921 & 0.6488 & \underline{0.6994} & \underline{0.3351} & \underline{0.3322} & 0.5656 & $-$0.0233 \\
\addlinespace
$\rho{=}2.0$ No Residual & 0.6849 & 0.6938 & \underline{0.7413} & \underline{0.7035} & \underline{0.3359} & \underline{0.3329} & \underline{0.5730} & -- \\
$\rho{=}2.0$ Residual & \underline{0.6931} & \underline{0.6997} & \underline{0.7452} & \underline{0.7044} & \underline{0.3369} & \underline{0.3369} & \underline{0.5861} & $+$0.0131 \\
\midrule
\textbf{GLU} &  &  &  &  &  &  &  &  \\
\textbf{Depth 2} &  &  &  &  &  &  &  &  \\
$\rho{=}1.0$ No Residual & \underline{0.6932} & 0.6991 & \underline{0.6875} & \underline{0.7013} & 0.3341 & 0.3345 & \underline{0.5714} & -- \\
$\rho{=}1.0$ Residual & 0.6881 & 0.6924 & 0.6816 & 0.6987 & 0.3333 & 0.3328 & 0.5447 & $-$0.0267 \\
\addlinespace
$\rho{=}2.0$ No Residual & \underline{0.6934} & 0.6965 & \underline{0.7108} & \underline{0.6951} & \underline{0.3365} & \underline{0.3388} & \underline{0.5747} & -- \\
$\rho{=}2.0$ Residual & \underline{0.6979} & 0.6987 & \underline{0.7151} & \underline{0.6990} & \underline{0.3373} & \underline{0.3401} & \underline{0.5813} & $+$0.0176 \\
\addlinespace
\textbf{Depth 3} &  &  &  &  &  &  &  &  \\
$\rho{=}1.0$ No Residual & 0.6660 & 0.6504 & 0.6659 & 0.6851 & 0.3317 & 0.3218 & 0.5600 & -- \\
$\rho{=}1.0$ Residual & 0.6642 & 0.6400 & 0.6601 & 0.6867 & 0.3313 & 0.3206 & 0.5505 & $-$0.0095 \\
\addlinespace
$\rho{=}2.0$ No Residual & \underline{0.7007} & \underline{0.6980} & \underline{0.7342} & \underline{0.7039} & \underline{0.3386} & \underline{0.3428} & \underline{0.5824} & -- \\
$\rho{=}2.0$ Residual & \underline{0.7042} & \underline{0.7025} & \underline{0.7371} & \underline{0.7056} & \underline{0.3392} & \underline{0.3448} & \underline{0.5889} & $+$0.0065 \\
\bottomrule
\end{tabular}
\end{table}

Next, we focus on the importance of upscaled representations within the intermediate layers. This projection is a common component of the modern Transformer feedforward block design~\citep{transformers}, with virtually all modern models adopting it, and has also been shown to improve the performance of even older architectures such as Recurrent Neural Networks~\cite{sharnn}. However, as demonstrated by the GLU results above, not all architectural modifications which improve Transformer networks appear to directly translate to improving our considerably-smaller network whose main purpose is dimensionality reduction.

Table~\ref{tab:depth_scale} presents the results of using an upscaled projection with a $\rho$ of 2, meaning that the intermediate representations' dimension is twice that of the inptu dimension, compared to a $\rho$ of 1 where intermediate representations are not upscaled. The overall results vary setting by setting, but ultimately appear to strongly favor upscaling. Interestingly, while it does not appear to considerably benefit GLU networks at a depth of 2, even resulting in a slight performance decrease, but mitigates large decreases in performance at the deeper depths of 3 and 4. For FFN blocks, results do show a similar preserving effect as depth increases, but a $\rho$ of 2 is superior to the no-upscaling setting across all model depths.

Overall, these results highlight that a higher-dimension intermediate dimension appear to contribute positively to stronger multi-vector retrieval performance, but also appear to have a stabilising effect, with the performance of similar model families using $\rho$=2 remaining more consistent across model depths while it greatly fluctuates without these upscaled representations.

\subsection{Residual Connections}
\label{sec:res_skipco}

Finally, we attempt to identify the effect of residual connections, and confirm their theoretical benefits of residual connections presented in~\ref{sec:skipco}. Table~\ref{tab:residual_comparison_full} presents a comparison of the effect of the use of residual connections on two different checkpoint families, across both their intermediate projection scale variants variants.

The results, presented in Table~\ref{tab:residual_comparison_full}, show an interesting phenomenon, which shines additional light on the seemingly stabilizing effect of larger intermediate projections highlighted above. Indeed, it appears that the use of residual connections consistently reduces the retrieval performance of models without upcasting. This effect appears milder in the simpler setting of the Depth 2 FFN block, our most simple model design, where the use of a residual projection has a negligible impact on performance, but is very noticeable in every other evaluated setting, resulting in large decreases.

On the other hand, when combined with a $\rho$ value of 2, where intermediate projections are upscaled, the use of residual connections significantly improve performance in all cases. Additionally, it seems to once again produce a stabilising effect, reducing the performance difference between various projection variants.

These results seem to the support the intuition expressed in Section~\ref{sec:skipco} in the sense that combining better projections with residual connections as part of these projections appear to result in greater performance, potentially as a result of better leveraging and "improving" the backbone model's projections rather than aggressively modifying them.

\section{Conclusion}

In this paper, we demonstrated the learning limitations imposed by the MaxSim operator of multi-vector retrieval models. We subsequently the hypothesis that these limitations are potentially harmful to downstream performance when combined with the simple, single-layer linear projection that is commonly used as the final layer of all existing multi-vector retrieval models.

We then proposed a series of improvements to the projection blocks of multi-vector models, discussing their potential benefits and limitations. Building on this proposal, we then trained numerous ColBERT models with all combinations of our proposed modifications.

Our results, evaluated across 5 independent training runs for each setting, demonstrate that the use of alternate projection heads appear to improve multi-vector performance across a variety of settings, with the best variant increasing performance by an average of over 2NDCG@10 points.

Finally, our exploration studies focus on independent modifications in order to better understand their role in this improved performance. We show that non-linearity, introduced either via GLU blocks or common activation functions, is not a significant performance driver, but that the use of modern FFN blocks with intermediate dimension upcasting and residual connections is crucial to our results.

While we propose theoretical explanations for these results, the learning process of Neural IR models, and particularly multi-vector models, is still poorly understand. We believe our empirical results are only an early step in the design of better multi-vector retrieval model architecture, and hope that they will support future work in better understanding their underlying mechanisms.

\appendix

\section{Compute Resources and Evaluation Choices}
\emph{Note: This appendix is currently placed before the bibliography to facilitate the review process.} \\

This study was performed using both RTX 4090 and NVidia A100 80GB GPUs, for evaluation. Each model training required an estimate 0.5 RTX 4090 hours and each full evaluation run 3 NVidia A100 hours. 

Due to the high potential cost of thousands of evaluation runs, not all checkpoints were fully evaluated on our full evaluation set, but rather evaluated using NanoBEIR\citep{nanobeir}, a downsampling of the BEIR evaluation suite which has been shown to be significantly correlated to full BEIR results. Subsequently, we confirmed that NanoBEIR results were highly correlated with our full evaluation results, both for the checkpoints whose results we report in Section~\ref{sec:results}, as well as for randomly selected checkpoints. This process allowed us to ensure that we were not missing any significant effect due to potential selection bias while studying the effects which we discuss in the main body.

\vskip 0.2in
\bibliography{sample}

\end{document}